\documentclass[fleqn]{article}
\usepackage{longtable}
\usepackage{graphicx}

\begin {document}
\begin{flushleft}
{\LARGE
{\bf Comment on ``Energy levels, oscillator strengths, and transition probabilities for sulfur-like scandium, Sc~VI" by El-Maaref et al.\,   [Indian J.  Phys. {\bf 91} 1029 (2017)]}
}\\

\vspace{1.5 cm}

{\bf {Kanti  M  ~Aggarwal}}\\ 

\vspace*{1.0cm}

Astrophysics Research Centre, School of Mathematics and Physics, Queen's University Belfast, \\Belfast BT7 1NN, Northern Ireland, UK\\ 
\vspace*{0.5 cm} 

e-mail: K.Aggarwal@qub.ac.uk \\

\vspace*{0.20cm}

Received: 13 May 2019; Accepted: 29 November 2019

\vspace*{1.0 cm}

{\bf Keywords:}  Energy levels, oscillator strengths, lifetimes, S-like scandium Sc~VI \\
\vspace*{0.5 cm}
{\bf PACS Nos.: } 32.70.Cs, 95.30.Ky
\vspace*{1.0 cm}

\hrule

\vspace{0.5 cm}

\end{flushleft}

\clearpage


\begin{abstract}

In this comment, through our independent calculations, we assess that the recently reported results of  El-Maaref et al.  [{\em Indian J.  Phys.}  {\bf 91} 1029 (2017)] for energy levels, oscillator strengths, radiative rates, and lifetimes are inaccurate and unreliable for several levels and transitions of S-like Sc~VI.

\end{abstract}

\clearpage

\section{Introduction}

Scandium (Sc) belongs to the group of rare earth elements and hence atomic data for its ions are useful for the studies of fusion plasmas \cite{kma1}. It is also an important element for the studies of astrophysical plasmas, as discussed by Pryce \cite{mhl}, and is a key element for understanding Am stars, as demonstrated by LeBlanc and Alecian \cite{am}. For this reason several workers have reported atomic data for its ions --  see for example, the early extensive work of Massacrier and Artru  \cite{ma}, who have reported atomic data for several ions of this element, i.e. Sc~III to Sc~XXI.

In one of the recent papers, El-Maaref et al. \cite{elm} have reported results  for energy levels, oscillator strengths (f-values), radiative rates (A-values), and lifetimes ($\tau$) for one of the important ions, namely S-like Sc~VI. For the calculations, they have adopted two independent atomic structure codes, namely the {\em configuration interaction version 3} (CIV3 \cite{civ3}) and the Los Alamos National Laboratory (LANL) code, which is mainly based on the relativistic Hartree-Fock code of Cowan \cite{cow}. Based on these two calculations, and comparisons made with the existing theoretical and experimental results, they have assessed their data to be accurate and useful for applications. However, we strongly disagree with their calculations and conclusions, as discussed below.

For most levels, the differences between the CIV3 and LANL energies are $\sim$2\% (or $\sim$ 8000~cm$^{-1}$ in absolute values), but up to 5\% for a few, such as of 3p$^4$~$^3$P$_{1,2}$ --  see table~2 of \cite{elm}. Since measurements of  energies have also been made for some levels of Sc~VI, these values have been compiled and assessed by the NIST (National Institute of Standards and Technology) team, and their recommended results are freely available at their website: {\tt https://physics.nist.gov/PhysRefData/ASD/levels\_form.html}. The measured energies are (generally) considered to be more accurate and therefore the {\em aim} of most calculations is to match their results with those. Therefore, as expected, El-Maaref et al. \cite{elm} have also made comparisons between their calculated energies and those listed by NIST. For lower levels, such as 1--9 (see table~2 of \cite{elm}), the CIV3 results are comparatively more accurate, but for the higher ones, such as 48--55, the LANL energies are in (much) better agreement with those of NIST. So there is no consistency. 

Unfortunately, experimental energies are not available for all the levels calculated by El-Maaref et al. \cite{elm}, and neither are the theoretical ones, as may be noted from their table~2. However, it is not strictly true because, as stated earlier, Massacrier and Artru  \cite{ma} have considered much larger number of levels, i.e. 1889, more than an order of magnitude than those calculated by El-Maaref et al., i.e. only 160. They could not make comparisons with the earlier work as perhaps they were unaware of it. Anyway, it is not our (great) concern at present about the number of levels or their accuracy. However, based on the comparisons made by El-Maaref et al. or the ones briefly discussed by us above, we can state with confidence that there is scope for improvement in the calculated energies for the levels of Sc~VI. But it may also be fair to state that there are no (large) discrepancies in their calculated energies. Therefore, we do not elaborate much on the energy levels and rather focus on larger discrepancies seen for other parameters, i.e. f- and A-values, and $\tau$ in their tables~3 and 4, respectively.

For the f-values (or related A-values in s$^{-1}$, because f$_{ij}$= 1.49$\times$10$^{-16} \lambda_{ji}^2 (\omega_j/\omega_i) A_{ji}$, where $\lambda$ is the transition wavelength in \AA\,  and $\omega$ is the statistical weight), discrepancies between the calculations of El-Maaref et al. \cite{elm} and those of Froese Fischer et al. \cite{mchf} with the {\em multi-configuration Hartree-Fock} (MCHF) method are up to three orders of magnitude for several transitions, see for example 33, 34 and 75 in their table~3. There are a few more calculations by other workers (see \cite{elm} for references), but we are not considering those because the ones by Froese Fischer et al. are perhaps the most accurate available to date. This is  because they have considered a very (very) large {\em configuration interaction} (CI), which is very important for moderately heavy ions of Sc. Similarly, differences between the two sets of $\tau$ values are up to three orders of magnitude, as may be noted for levels 35--37 in table~4 of \cite{elm}. Since the results of El-Maaref et al. are more recent, these are also `expected' to be more accurate. Unfortunately that is not the case. More importantly, since these are the only two calculations available for $\tau$ for the levels of Sc~VI, such large differences cannot be ignored but rather need to be explained, resolved and understood, which has not been done by El-Maaref et al. Therefore, through our independent calculations we assess which data are more accurate and why.

\section {Calculations}

For the  determination of energy levels, radiative rates,  oscillator strengths, and lifetimes,  we have adopted  the {\em General-purpose Relativistic Atomic Structure Package} (GRASP). The original code, developed  by Grant et al. \cite{grasp}, has undergone numerous modifications and improvements by many workers in the past, and as a result of it  several published versions are available in the literature, but the one employed by us is known as GRASP0 and is currently hosted at the website {\tt http://amdpp.phys.strath.ac.uk/UK\_APAP/codes.html}. This is a fully {\em relativistic} code based on the $jj$ coupling scheme, although the two-body relativistic operators included in this (but not in CIV3) are not too important for a moderately heavy ion Sc~VI. Therefore, differences (if any) in energies (or other related parameters) with other calculations will not be because of this, but for other reasons. Finally, in the calculations we have preferred the choice of {\em extended average level} (EAL)  in which a weighted (proportional to 2j+1) trace of the Hamiltonian matrix is minimised. This option produces a compromise set of orbitals describing closely lying states with moderate accuracy, and results obtained  with other options, such as average level (AL), are comparable for all levels.

El-Maaref  et al. \cite{elm} have considered 160 levels belonging to 8 configurations, namely 3s$^2$3p$^4$, 3s3p$^5$, 3s$^2$3p$^3$3d, 3s$^2$3p$^3$4s/4p/4d, and 3s$^2$3p$^3$5s/5p. In the calculations they have included {\em limited}\, CI, mainly among these 8 configurations. We have performed several calculations, gradually increasing the number of orbitals and configurations, and subsequently the CI. Our final calculations include 4498 levels, which belong to 41 configurations, namely 3s$^2$3p$^4$, 3s3p$^5$, 3p$^6$, 3s$^2$3p$^3$3d, 3s3p$^4$3d, 3p$^5$3d, 3s$^2$3p$^2$3d$^2$, 3s3p$^3$3d$^2$, 3s$^2$3p$^3$4$\ell$, 3s3p$^4$4$\ell$, 3s$^2$3p$^2$3d4$\ell$, 3s$^2$3p3d$^3$, 3s3p$^3$3d4$\ell$, 3p$^4$3d$^2$, 3p$^5$4$\ell$, 3s3p$^2$3d$^3$, 3s$^2$3p$^3$5$\ell$, and 3s$^2$3p$^3$6$\ell$ ($\ell \le$ g). Inclusion of such a large CI is necessary for two reasons: (i) Sc~VI is only moderately heavy and, more importantly, (ii) levels of these configurations {\em intermix} -- see for example table~5 of Massacrier and Artru  \cite{ma}. El-Maaref  et al. have ignored many of the important configurations, such as 3s3p$^4$3d whose 56 levels appear early on, well before those of 3p$^3$4$\ell$. Another example is 3p$^6$, which generates only one level but intermixes with those of 3p$^3$3d. El-Maaref  et al. have mostly included those configurations whose energy levels appear on the NIST website. Inclusion of these additional configurations not only improves the accuracy of the calculated energies but also (significantly) affects the further determinations of $\tau$, because contributions arising from these {\em missing} levels are also accounted for. Conversely, the 160 levels considered by them are {\em not} the lowest. To calculate lifetime for the highest level of their calculations, i.e. 3p$^3$5p~$^1$S$_0$, one needs to consider at least 524 levels!

\section {Oscillator strengths}

As stated in section~1, our focus is on the results for f- (or A-) values. For weaker transitions (say with f $<$ 0.1), differences among different calculations may sometimes be considerably large, particularly for those with f $<$ 10$^{-3}$. This is because their resultant matrices are highly sensitive to the {\em small} mixing coefficients from different levels and configurations. However, comparatively stronger transitions are generally immune to the additive or cancellation effects of the coefficients, because of their larger magnitudes. Therefore, it is much easier to establish the (in)accuracy of a calculation for such transitions by comparing their f-values.

In Table~1 we list a few transitions which are comparatively strong(er) with f-values from the CIV3 calculations of El-Maaref  et al. \cite{elm}, Froese Fischer et al. \cite{mchf}  with MCHF, and our own work with GRASP. These representative transitions are sufficient to highlight the similarity (and or the discrepancies) among different calculations. For all transitions listed here (and many more -- see table~A of \cite{adndt}) there is a satisfactory agreement between the MCHF and GRASP results, but the CIV3\,  f-values of El-Maaref  et al.  differ by up to three orders of magnitude. Clearly, the results of  El-Maaref  et al. are neither correct nor reliable.

\begin{table}
\caption{Comparison of oscillator strengths (f-values) for some transitions of Sc~VI. a$\pm$b $\equiv$ a $\times$ 10$^{{\pm}\rm b}$.} 
\begin{tabular}{lllllll} \hline
\\        
Lower    Level              &   Upper Level                                 &   CIV3    &  MCHF  &  GRASP    \\ 
\hline \\												       
  3p$^4$~$^3$P$_2$  &  3p$^3$($^2$P)3d~$^3$P$^o_2$  &   0.148  &   0.018  &   0.016   \\
  3p$^4$~$^3$P$_2$  &  3p$^3$($^2$D)3d~$^3$P$^o_2$  &   0.006  &   0.792  &   0.771   \\
  3p$^4$~$^3$P$_2$  &  3p$^3$($^2$D)3d~$^3$P$^o_1$  &   6.7-4  &   0.183  &   0.161   \\
  3p$^4$~$^3$P$_2$  &  3p$^3$($^4$S)3d~$^3$D$^o_3$  &   0.589  &   1.480  &   1.490   \\
  3p$^4$~$^3$P$_2$  &  3p$^3$($^4$S)3d~$^3$D$^o_2$  &   0.106  &   0.231  &   0.238   \\
  3p$^4$~$^3$P$_1$  &  3p$^3$($^4$S)3d~$^3$D$^o_2$  &   0.178  &   1.360  &   1.364   \\
  3p$^4$~$^3$P$_0$  &  3s3p$^5$~$^3$P$^o_1$	    &   0.102  &   0.056  &   0.052   \\
  3p$^4$~$^3$P$_0$  &  3p$^3$($^2$D)3d~$^3$P$^o_1$  &   0.003  &   0.831 &   0.953   \\
\\ \hline
\end{tabular} 

\begin{flushleft}
{\small
CIV3: earlier calculations of El-Maaref et al. \cite{elm} with  the {\sc civ3} code  \\
MCHF: earlier calculations of Froese Fischer et al. \cite{mchf} with  the {\sc mchf} code  \\
GRASP: present calculations with  the {\sc grasp} code  \\
}
\end{flushleft}
\end{table}

\section {Lifetimes}

The lifetime of a level is determined as $\tau$ (s) = 1.0/$\Sigma_{i}$A$_{ji}$ where the summation runs over all types of transitions, although the electric dipole (E1) ones  are normally the most dominant in magnitude, and hence  more important. Nevertheless, we have included contributions from other types as well, namely magnetic dipole (M1), electric quadrupole (E2), and magnetic quadrupole (M2), which have been ignored by El-Maaref  et al. \cite{elm}, but have also been considered by Froese Fischer et al. \cite{mchf}, as expected.

In Table~2 we list only those levels for which differences between the CIV3 and MCHF values of $\tau$ are very large, i.e. up to three orders of magnitude. Furthermore,  for brevity only the lowest few levels are considered here. Agreement between the GRASP and MCHF calculations is generally highly satisfactory and is within a factor of two. These differences are understandable because f-values for most of these levels are rather small in magnitude. However, discrepancies with the corresponding CIV3 results of El-Maaref  et al. \cite{elm} are indeed very large, and have arisen because of the corresponding differences in the f- (and A-) values, noted in section~3. 

\begin{table}
\caption{Comparison of lifetimes ($\tau$, s) for some levels of Sc~VI. a$\pm$b $\equiv$ a $\times$ 10$^{{\pm}\rm b}$.} 
\begin{tabular}{lllllll} \hline
\\        
Level                                         &   CIV3    &  MCHF  &  GRASP    \\ 
\hline \\												       
3p$^3$($^2$P)3d~$^3$P$^o_0$  &  2.25-10  &  2.50-09  &  1.59-09  \\
3p$^3$($^2$P)3d~$^3$P$^o_1$  &  1.51-10  &  1.86-09  &  8.91-10  \\
3p$^3$($^2$P)3d~$^3$P$^o_2$  &  8.02-11  &  8.66-10  &  5.49-10  \\
3p$^3$($^2$D)3d~$^3$P$^o_2$  &  1.71-09  &  1.28-11  &  1.25-11  \\
3p$^3$($^2$D)3d~$^3$P$^o_1$  &  5.12-09  &  1.29-11  &  1.26-11  \\
3p$^3$($^2$D)3d~$^3$P$^o_0$  &  1.05-08  &  1.25-11  &  1.22-11  \\
\\ \hline
\end{tabular} 

\begin{flushleft}
{\small
CIV3: earlier calculations of El-Maaref et al. \cite{elm} with  the {\sc civ3} code  \\
MCHF: earlier calculations of Froese Fischer et al. \cite{mchf} with  the {\sc mchf} code  \\
GRASP: present calculations with  the {\sc grasp} code  \\
}
\end{flushleft}
\end{table}

\section{Conclusions}

In a  recent paper El-Maaref et al. \cite{elm} have reported results for energy levels, f-values, A-values, and $\tau$ for the levels/transitions of S-like Sc~VI. Most of their results are based on the CIV3 calculations. While their calculated energies are (nearly) acceptable for most levels, the corresponding results for other parameters are not, because these differ from the earlier available as well as our present work for several transitions, by up to three orders of magnitude. The main reason for such large discrepancies is the inclusion of limited (and arbitrary) CI  in their work. Much more accurate, and larger, calculations already exist in the literature, which are by  Froese Fischer et al. \cite{mchf} and Massacrier and Artru  \cite{ma}, respectively. However, there is still scope for further improvement as well as extension over  the existing results. This is because levels of Froese Fischer et al. are confined to those of the 3s$^2$3p$^4$, 3s3p$^5$ and 3s$^2$3p$^3$3d configurations only. Massacrier and Artru, on the other hand, have included considerably larger CI (among 51 configurations) and have reported data for 1889 levels, but their results for A-values are confined to the E1 transitions alone, whereas those with other types (E2, M1 and M2) are also desirable in the modelling of plasmas, and particularly in the determination of lifetimes. Therefore, in a recent paper \cite{adndt}  we have reported a complete set of data for all levels and their transitions, not only for S-like Sc~VI, but also for V~VIII, Cr ~X, and Mn~X

With the easy access to atomic structure (and scattering) codes, it has become (rather) straightforward to generate atomic data for various parameters. However, such data may not be of any use and may often be confusing and misleading, because generating accurate and reliable data, which can be confidently applied for modelling or diagnostics of plasmas,  is still difficult to produce as several checks and comparisons are required, particularly when the earlier available results are in paucity. This is the main reason that large discrepancies, of orders of magnitude, are often observed in the reported data (for any ion), as noted here for Sc~VI and highlighted earlier for many others  in our earlier publications \cite{fst}, \cite{atoms}. Finally, for the benefit of the readers we will like to note that the earlier reported results by El-Maaref and co-workers are in large errors for other ions also, as discussed and explained by us for S-like Mn~X \cite{mnx},  Kr-like W~XXXIX \cite{w39} and Zn-like W~XLV~\cite{w45}.


\end{document}